\documentclass[aps,pra,amsmath,amsfonts,amssymb,twocolumn,superscriptaddress,showpacs]{revtex4-1}
\usepackage{amssymb,amsmath,amsfonts}
\usepackage{graphicx,psfrag,color}
\usepackage{dcolumn}
\usepackage{bbm}
\usepackage{mathbbol}
\usepackage{braket}
\usepackage{times}
\usepackage{graphicx}
\usepackage{verbatim}
\usepackage{hyperref}
\usepackage{subcaption}
\usepackage{ragged2e}
\usepackage[normalem]{ulem}

\newcommand*{\balancecolsandclearpage}{%
  \close@column@grid
  \clearpage
  \twocolumngrid
}

\begin{document}
\flushbottom

\title{Absorption-based qubit estimation in discrete-time quantum walks}

\author{Edgard P. M. Amorim}
\email{edgard.amorim@udesc.br} 
\affiliation{Departamento de F\'isica, Universidade do Estado de Santa Catarina, 89219-710 Joinville, SC, Brazil}

\author{Lorena R. Cerutti}
\affiliation{Departamento de F\'isica, Universidade do Estado de Santa Catarina, 89219-710 Joinville, SC, Brazil}

\author{O. P. de S\'a Neto}
\affiliation{Coordenação de Ci\^encias da Computa\c{c}\~ao, Universidade Estadual do Piau\'i, 64202–220 Parna\'iba, PI, Brazil}

\author{M. C. de Oliveira}
\email{marcos@ifi.unicamp.br}
\affiliation{Instituto de F\'isica  Gleb Wataghin, Universidade Estadual de Campinas, 13083-859, Campinas, SP, Brazil}
\date{\today}

\begin{abstract}
We investigate state estimation in discrete-time quantum walks with a single absorbing boundary. Using a spectral approach, we obtain closed expressions for the escape probability as a function of the initial coin state and the boundary position, together with the corresponding classical Fisher information for a binary absorption readout. Comparison with the single-copy quantum Fisher information reveals a clear complementarity: near boundaries carry broad information about the polar (Bloch-sphere) angle of the coin state, whereas moderate or distant boundaries reveal phase-sensitive regions. Because a single boundary probes only one information direction, combining two boundary placements yields, generically, a full-rank Fisher matrix and tight joint Cramér-Rao bounds while retaining a binary measurement without mode-resolved tomography. We also discuss a restricted-readout photonic implementation in which an on-chip sink realizes the absorber, and we frame the resulting advantage as a potential reduction in measurement-setting and reconfiguration overhead for low-dimensional parameter estimation tasks in architectures where direct projective access to the coin is unavailable. Our results show that absorption in quantum walks defines an analytically tractable restricted-access primitive for coin-state estimation.
\end{abstract}


\maketitle

\section{Introduction}\label{sec.1}

Quantum walks are the quantum analogs of classical random walks, describing the coherent evolution of a quantum particle over a discrete or continuous set of positions \cite{aharonov1993quantum}. In the discrete-time version, the walker possesses an internal degree of freedom—often called a coin space—that governs the direction of motion on the external position space. The total state of the system is therefore a vector in the tensor product of the coin and position Hilbert spaces. At each time step, the state evolves unitarily through the sequential action of two operators: a quantum coin operator, which performs a rotation in the coin subspace (analogous to tossing a coin in the classical walk), and a conditional shift operator, which displaces the walker according to the coin state. The combination of these operations leads to interference between different paths, producing probability distributions that differ markedly from their classical counterparts \cite{pearson1905random}. These interference effects give rise to distinctly quantum phenomena such as ballistic spreading, localization, and entanglement between the coin and position degrees of freedom, which make quantum walks a valuable framework for studying quantum transport, simulation, and algorithmic speedups \cite{kempe2003quantum,venegas2012quantum,portugal2013quantum}.

Quantum walks have been extensively employed as the underlying mechanism for quantum search algorithms \cite{shenvi2003quantum,tulsi2008faster}, and they constitute a universal model for quantum computation, capable of implementing a complete set of logical gates \cite{childs2009universal,lovett2010universal}. They also exhibit rich dynamical behavior: depending on the initial state and the sequence of quantum coins throughout the walk, they can display localization~\cite{vieira2014entangling,da2021localization}, diffusive spreading with maximal entanglement between coin and position states~\cite{vieira2013dynamically}, and solitonlike dynamics enabling high-fidelity state transfer~\cite{orthey2019connecting,ghizoni2019trojan,vieira2021quantum,engster2024high}. Several physical platforms have been proposed and employed for their realization, including photonic devices, trapped ions, and superconducting circuits \cite{Silberhorn18, Solano2010,Lozada-Vera2016,PhysRevB.95.144506, wang2013physical}.

A practical limitation in several quantum-walk platforms is that the coin degree of freedom (polarization, internal level, or other ancilla) may not be directly measurable in arbitrary bases at the output, or doing so may require extensive reconfiguration and phase-stable interferometry. In integrated photonics, for instance, controlled loss channels or ``sinks'' can be engineered on-chip, while mode-resolved polarization/state tomography across many paths can be costly in settings and stability requirements. This motivates an estimation-theoretic question that differs from standard qubit metrology—If the only accessible measurement is a global binary outcome (absorbed versus escaped), how much information about the initial coin qubit can still be inferred, and how should one place absorbers to maximize it?

Classical and quantum walkers behave differently in the presence of an absorbing boundary. Consider a simple symmetric \emph{classical} random walk on $x\in\mathbb{Z}$ with an absorbing site placed one step from the starting position (e.g., start at \(x=0\), absorb at \(x=1\)). With hop probability \(p=1/2\) to the left/right, the one-dimensional walk is \emph{recurrent}: The hitting probability of any given site is 1. Consequently, the eventual absorption probability is \(P_A = 1\) and the escape (survival) probability is $P_E=\lim_{n\rightarrow\infty} 1/2^n=0$ (see, e.g., the recurrence discussion in Ref. \cite{kempe2003quantum}). In contrast, for a discrete-time quantum walk, the total absorption probability depends on the coin operator and the initial coin state. Interference generically leaves a nonzero amplitude on trajectories that never reach the boundary, yielding \(P_E = 1 - P_A > 0\) even when the absorbing site is placed a finite distance from the initial position. This strict separation from the classical case—nonunit absorption on the half-line—has been analyzed in detail (see Refs. \cite{bach2004onedimensional,konno2003absorption}).

A substantial literature exists on discrete-time quantum walks with absorbing boundaries. Different authors have employed distinct analytical techniques to obtain absorption (hitting) probabilities. Bach \emph{et al.} derived closed-form expressions for walks with one and two static absorbing boundaries using (1) a combinatorial approach based on generating functions and (2) an eigenfunction (spectral) method built on the Fourier transform of the walk dynamics \cite{bach2004onedimensional,bach2009absorption}. Independently, Konno \emph{et al.} obtained equivalent results via a path-counting approach grounded in a Feynman path-sum formulation \cite{konno2003absorption}. Extensions include moving absorbing boundaries \cite{kwek2011onedimensional} and position-dependent (partially) reflecting boundaries \cite{wang2017quantum}. While absorption probabilities themselves are well known, their information-geometric structure under a binary absorption/escape readout and its dependence on boundary placement—particularly the resulting rank deficiency and complementarity in the Fisher matrix—have not, to our knowledge, been analyzed in this estimation-theoretic form.

Our proposal is to address the following inference question: assuming that we do not have access to the initial state of the walker, how much information do the (measurable) absorption/escape probabilities provide about that state? We adopt an estimation-theoretic point of view and quantify informativeness of the absorption readout about the state parameters via the (classical) Fisher information \cite{paris2004quantum,paris2009quantum}. This type of binary measurement is particularly appealing in architectures where internal degrees of freedom or individual output modes are not experimentally resolved. In that setting, one can still probe the underlying coin qubit by repeating the walk many times and recording only whether the walker is absorbed or escapes. Throughout this work, we adopt a per-trial (``single-copy'') metrological viewpoint: each run of the walk encodes one copy of the unknown coin state, and many independent repetitions are used to estimate its Bloch angles via Fisher information and Cramér-Rao bounds. The present strategy is particularly relevant for integrated-photonic implementations using an on-chip sink, since it may offer a reduced number of measurement configurations for low-dimensional parameter estimation in architectures where direct, mode-resolved tomography is impractical. For that end, we derive an analytical expression for the escape probability of a discrete-time quantum walk with an absorbing boundary, expressed as a function of the initial coin state (qubit) and the boundary position, using the spectral (eigenfunction) method in quasimomentum space. Consequently, we obtain analytical Fisher information for the coin parameters, benchmarked against the single-copy quantum Fisher information (QFI). 

The paper is organized as follows. Section~\ref{sec.2} summarizes the quantum-walk formalism. Section~\ref{sec.3}, together with Appendixes~\ref{append:1} and \ref{append:2}, reviews and adapts the eigenfunction approach of Ref.~\cite{bach2004onedimensional} to obtain analytical expressions for the escape probability as a function of the coin-state parameters and boundary position. Section~\ref{sec.4} develops the coin-state estimation framework and efficiencies. Section V discusses the relevance of the present findings for integrated-photonic platforms, together with practical design considerations and a photonics outlook. Finally, Sec.~\ref{sec.6} presents the main conclusions.

\section{Quantum walks}\label{sec.2}

We consider a discrete-time quantum walk on the one-dimensional lattice. The quantum walker state belongs to the total Hilbert space \(\mathcal{H}=\mathcal{H}_c\otimes\mathcal{H}_p\), where the two-dimensional coin space
\(\mathcal{H}_c=\mathrm{span}\{\ket{L},\ket{R}\}\) encodes an internal qubit and the position space
\(\mathcal{H}_p=\mathrm{span}\{\ket{j}: j\in\mathbb{Z}\}\) encodes lattice sites, with \(\braket{j|j'}=\delta_{j,j'}\). The walker is initialized at the origin with a general coin state; therefore, the joint initial state is 
\begin{equation}
\ket{\Psi(0)}=\left[L(0,0)\ket{L}+R(0,0)\ket{R}\right]\otimes\ket{0},
\label{Psi0}
\end{equation} 
where $L(0,0)=\cos(\alpha/2)$ and $R(0,0)=e^{i\beta}\sin(\alpha/2)$ are the initial amplitudes with $\alpha\in[0,\pi]$ and $\beta\in[0,2\pi]$ being the usual polar and azimuth angles in the Bloch-sphere representation \cite{nielsen2010quantum}. The state evolves by $\ket{\Psi(t)}=U\ket{\Psi(t-1)}$, implying that after $t$ time steps $\ket{\Psi(t)}=U^t\ket{\Psi(0)}$ for $U$ constant over all the time and positions. The time-evolution operator, $U=S(C\otimes\mathbbm{1}_p)$, includes the biased Hadamard quantum coin,
\begin{equation}
\displaystyle
C=
\begin{bmatrix}
\sqrt{\rho}    &   \sqrt{1-\rho} \\
\sqrt{1-\rho}  &  -\sqrt{\rho}
\end{bmatrix},
\label{Hadamard}
\end{equation}
with a bias parameter $\rho\in[0,1]$ (unbiased for $\rho=1/2$), where  $\mathbbm{1}_p$ is the identity in the walker Hilbert space. It also includes the conditional displacement operator $S$, such that
\begin{align}
S(\ket{L}\otimes\ket{j})&=\ket{L}\otimes\ket{j-1}, \nonumber \\
S(\ket{R}\otimes\ket{j})&=\ket{R}\otimes\ket{j+1},
\label{SL&SR}
\end{align}
which moves the $\ket{L}$ ($\ket{R}$) amplitude to the left (right) neighboring lattice position. 

To evaluate the escape probability of a quantum walker, we should take the asymptotic limit $t\rightarrow\infty$ of the probability over the positions out of the absorption barrier. However, while the step operator is not diagonal in the position basis, it becomes diagonal in quasimomentum, so it is convenient to change the basis from $\mathcal{H}_p$ to the dual $k$ space $\mathcal{H}_k$ to perform a spectral decomposition (see Appendix~\ref{append:1} for more details). 

\section{Absorption barrier}\label{sec.3}

\begin{figure*}[ht]
\includegraphics[width=0.8\linewidth]{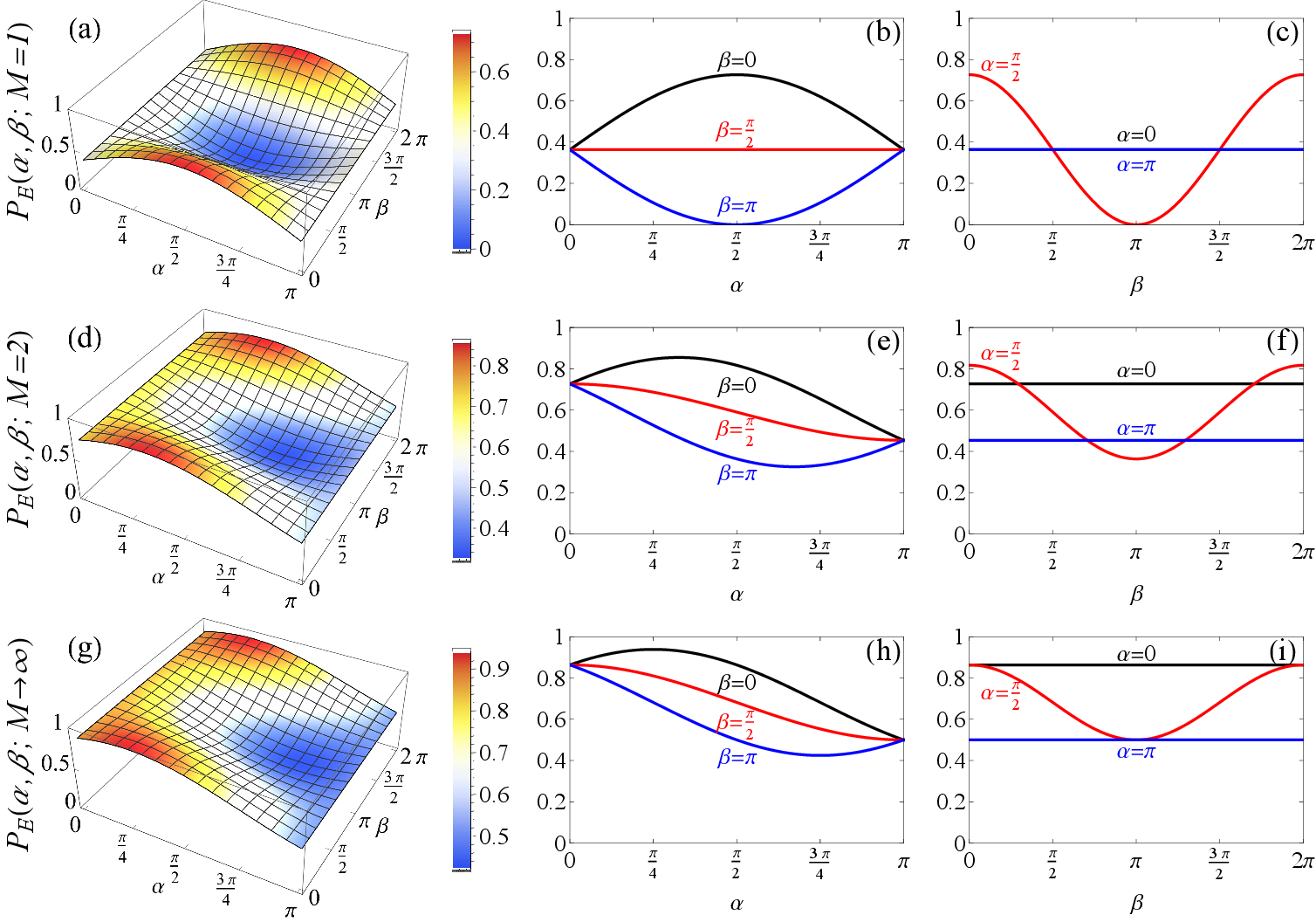}
\caption{\justifying Escape probability $P_E(\alpha,\beta;M)$ as a function of the Bloch angles $(\alpha,\beta)$ for barrier positions (a)–(c) $M=1$, (d)–(f) $M=2$, and (g)–(i) $M\to\infty$. The first column shows the surface $P_E(\alpha,\beta;M)$ for each case, while the second and third columns display one-dimensional cuts of these surfaces for $\alpha,\beta = 0$ (black), $\pi/2$ (red), and $\pi$ (blue). Note that in panel (c) the curves for $\alpha=0$ and $\alpha=\pi$ coincide.}
\label{fig.2}
\end{figure*}

Let us now introduce the recurrence equations considering an absorption barrier placed on $j=M$, where $M>0$. Note that for the positions $j<M-1$, we have the recurrence equations for a biased Hadamard walk with no barrier influence,
\begin{equation}
\begin{bmatrix}
L(j,t)\\
R(j,t)\\
\end{bmatrix}
\!=\! 
\begin{bmatrix}
\!\sqrt{\rho} L(j\!+\!1,\!t\!-\!1)\!+\!\sqrt{1\!-\!\rho} R(j\!+\!1,\!t\!-\!1)\\
\!\sqrt{1\!-\!\rho} L(j\!-\!1,\!t\!-\!1)\!-\!\sqrt{\rho} R(j\!-\!1,\!t\!-\!1)\\
\end{bmatrix}
\!.
\label{REHad}
\end{equation}
For $j\geq M$, as our walker starts on $j=0$, the amplitudes arrive at the barrier only coming from left to right. After the amplitudes cross the barrier, they do not return to the opposite direction. These amplitudes represent the part of the state absorbed by the barrier. Therefore, the barrier behaves like an identity operator,
\begin{equation}
\begin{bmatrix}
L(j,t)\\
R(j,t)\\
\end{bmatrix}
\!=\! 
\begin{bmatrix}
L(j\!+\!1,t\!-\!1)\\
R(j\!-\!1,t\!-\!1)\\
\end{bmatrix}
\!.
\label{REId}
\end{equation}
At last, for $j=M-1$, note that the amplitude $L(M,t-1)$ originates from the barrier at the right, while the amplitude $R(M-1,t)$ comes from the Hadamard coin at the left; therefore,
\begin{equation}
\begin{bmatrix}
L(M\!-\!1,t)\\
R(M\!-\!1,t)\\
\end{bmatrix}
\!=\!
\begin{bmatrix}
L(M,t\!\!-\!\!1)\\
\!\sqrt{1\!\!-\!\!\rho}L(M\!\!-\!\!2,t\!\!-\!\!1)\!-\!\sqrt{\rho}R(M\!\!-\!\!2,t\!\!-\!\!1)\!\\
\end{bmatrix}\!.
\label{REIdHad}
\end{equation}

The escape probability $P_E$ is obtained by imposing an absorbing boundary at site $j=M-1$ and solving the walk dynamics via a spectral decomposition (see Appendix~\ref{append:1}) combined with the method of images (see Appendix~\ref{append:2}). In this construction, a mirror walker is introduced symmetrically with respect to the boundary so that the superposition of real and mirror contributions enforces $L(M-1,t)=0$ at all times. The resulting eigenmode expansion yields closed expressions for the left-moving components that survive in the long-time limit, from which one derives integral formulas for $P_E(M)$ assuming a Hadamard walk ($\rho=1/2$) \cite{bach2004onedimensional}. All these manipulations drive us to the following closed expressions of the escape probability as a function of the initial coin state:
\begin{align}
P_E(\alpha,\beta)&=\xi_1\cos^2\left({\frac{\alpha}{2}}\right)+\xi_2\sin^2\left({\frac{\alpha}{2}}\right)\nonumber \\
                 &+\xi_3\cos\left({\frac{\alpha}{2}}\right)\sin\left({\frac{\alpha}{2}}\right)\cos\beta,
\label{PEM} 
\end{align}
where the constants $\xi_i$ are listed in Table~\ref{tab.1} for selected values of $M$. Figure~\ref{fig.2} displays the escape probability $P_E(\alpha,\beta;M)$ obtained from Eq.~\eqref{PEM} using the coefficients $\xi_i$ from Table~\ref{tab.1} for barrier positions $M=1$, $M=2$, and $M\to\infty$.
\begin{table}[b]
\caption{Calculated values for $\xi_1$, $\xi_2$, and $\xi_3$.}
\label{tab.1}
\begin{ruledtabular}
\begin{tabular}{lccc} 
\hline
$M$      & $\xi_1$                     & $\xi_2$                   & $\xi_3$                   \\
\hline
1        & $1-\frac{2}{\pi}$           & $1-\frac{2}{\pi}$         & $2-\frac{4}{\pi}$         \\[+2pt]
2        & $2-\frac{4}{\pi}$           & $3-\frac{8}{\pi}$         & $3-\frac{8}{\pi}$         \\[+2pt]
3        & $4-\frac{10}{\pi}$          & $13-\frac{118}{3\pi}$     & $11-\frac{100}{3\pi}$     \\[+2pt]
4        & $14-\frac{124}{3\pi}$       & $65-\frac{608}{3\pi}$     & $53-\frac{496}{3\pi}$     \\[+2pt]
5        & $66-\frac{614}{3\pi}$       & $341-\frac{16~046}{15\pi}$ & $277-\frac{13~036}{15\pi}$ \\[+2pt]
$\infty$ & $\frac{3}{2}-\frac{2}{\pi}$ & $\frac{1}{2}$             & $1-\frac{2}{\pi}$         \\[+2pt] 
\hline
\end{tabular}
\end{ruledtabular}
\end{table}
For $M=1$, the escape probability has a symmetric profile, as shown in the first row of Fig.~\ref{fig.2} [panels (a)–(c)]. This indicates that qubits on the northern hemisphere of the Bloch sphere have the same escape probability as the corresponding qubits on the southern hemisphere. Moreover, only for $M=1$ does the escape probability reduce to
\begin{equation}
P_E(\alpha,\beta;M\!=\!1)=\left(1-\frac{2}{\pi}\right)\bigl(1+\sin\alpha\cos\beta\bigr),
\label{PE1} 
\end{equation}
which vanishes at $(\alpha,\beta)=(\pi/2,\pi)$. A Hadamard coin applied to this particular qubit at site $j=0$ yields the state $\ket{R}$, and the conditional displacement operator moves it to the right, where it is completely absorbed by the barrier at $j=1$. For $M>1$, the escape probability becomes asymmetric: qubits with a larger amplitude of $\ket{L}$ have a higher escape probability than those with a larger amplitude of $\ket{R}$, and this difference becomes more pronounced as $M$ increases. This behavior reflects the fact that the absorbing barrier is placed at a positive position $j>0$. These configurations, corresponding to $M=1$ (nearest boundary), $M=2$ (moderate distance), and $M\to\infty$ (larger distance), will serve as our starting point for estimating the qubit parameters via the Fisher information in the next section.

\section{Coin-State Estimation}\label{sec.4}

In natural systems such as quantum walks, we have intrinsic statistical properties influenced by observable and unobservable physical quantities. The observable physical quantities are all those that can be measured. In contrast, the unobservable ones influence the curvature of the probability of detection in the system, and they are defined as parameters. We view absorption readout as a binary measurement with success probability $P_E(\alpha,\beta;M)$ given by Eq.~(\ref{PEM}). For $N$ independent runs of the walk at a fixed boundary position $M$, the number of escapes $k$ follows a binomial law with likelihood
\begin{equation}
\mathcal{L}(\alpha,\beta)\propto
P_E(\alpha,\beta;M)^{\,k}\,[1-P_E(\alpha,\beta;M)]^{\,N-k}.
\end{equation}
We can estimate the initial state of a quantum walk constructed inside a box by measuring $P_E(\alpha,\beta;M)$. 
For a single parameter $\theta\in\{\alpha,\beta\}$, the per-trial (classical) Fisher information $F_{\theta}(\alpha,\beta;M)\equiv F_{\theta}(\alpha,\beta)$ \cite{Fisher} is given by 
\begin{equation}
F_{\theta}(\alpha,\beta)=
\frac{\big[\partial_{\theta}P_E(\alpha,\beta;M)\big]^2}
{P_E(\alpha,\beta;M)\,[1-P_E(\alpha,\beta;M)]}.
\label{eq:scalarFI}
\end{equation}
Because the readout is a single classical binary variable (escape versus absorption), the classical Fisher information fully characterizes the achievable local precision for any unbiased estimator via the Cramér-Rao bound, making it the natural figure of merit in the restricted-measurement scenario considered here. With $N$ i.i.d.\ trials, the Cramér-Rao bound \cite{c1,c2,R} for any unbiased estimator $\theta$ reads
\begin{equation}
\mathrm{Var}(\theta)\ \ge\ \frac{1}{N\,F_{\theta}(\alpha,\beta)},
\end{equation}
showing that a larger $F_{\theta}(\alpha,\beta)$ yields a tighter (smaller) Cramér-Rao lower bound for $\mathrm{Var}({\theta})$, and therefore allowing, in principle, more precise estimation of ${\theta}$.
Using Eq.~\eqref{PEM}, we obtain
\begin{equation}\label{RF}
F_\alpha(\alpha,\beta)=\frac{\left[(\xi_{2}\!-\!\xi_{1})\sin\alpha\!+\!\xi_3\cos\alpha\cos\beta\right]^2}
{f(\alpha,\beta)\left[2\!-\!f(\alpha,\beta)\right]},
\end{equation} 
and
\begin{equation}
F_{\beta}(\alpha,\beta)=\frac{\left[\xi_{3}\sin\alpha\sin\beta\right]^2}{f(\alpha,\beta)\left[2\!-\!f(\alpha,\beta)\right]},
\end{equation}
with $f(\alpha,\beta)=\xi_2\!+\!\xi_1\!-\!(\xi_2\!-\!\xi_1)\cos\alpha\!+\!\xi_3\sin\alpha\cos\beta$, where $\xi_i$ depend on $M$ as seen in Table~\ref{tab.1}.

\begin{figure}[h]
\includegraphics[width=\linewidth]{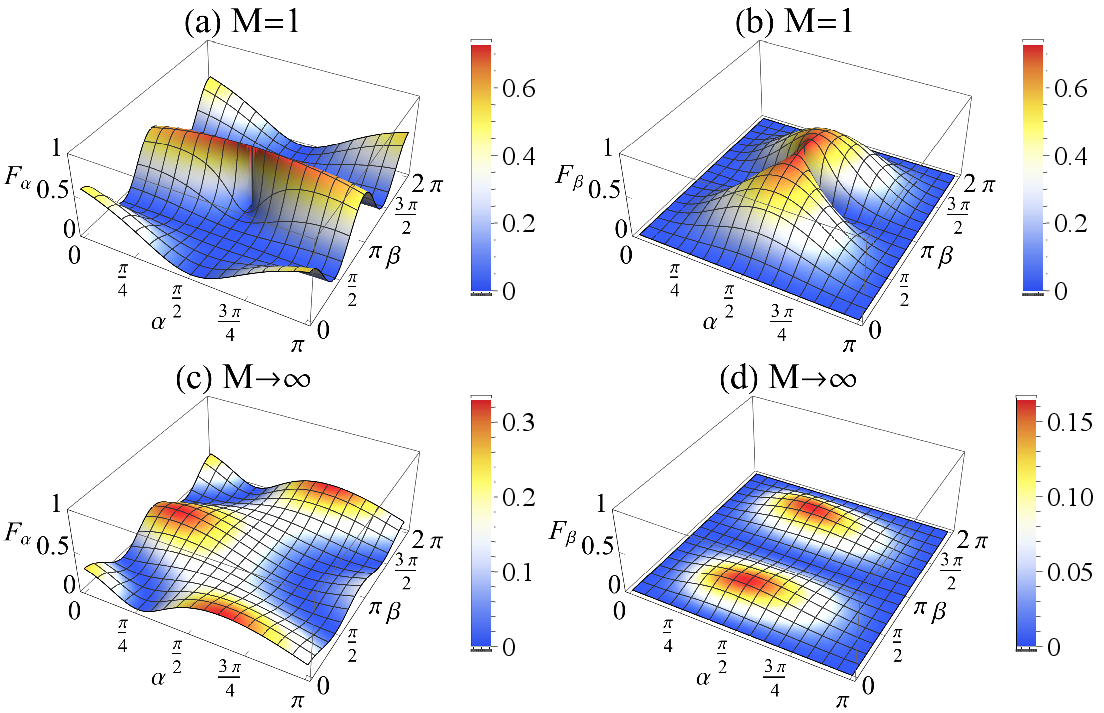}
\caption{\justifying Fisher information $F_\alpha$ (left column) and $F_\beta$ (right column) as functions of the initial qubit ($\alpha$,$\beta$) for the barrier positions (a), (b) $M=1$ and (c), (d) $M\rightarrow\infty$.}
\label{fig.3}
\end{figure}

Figure~\ref{fig.3} shows the Fisher information $F_\alpha$ and $F_\beta$ as functions of the Bloch angles $(\alpha,\beta)$ of the initial qubit. For $M=1$, we have an outstanding estimate for $\alpha$ when $\beta=\pi$, except at $\alpha=\pi/2$. This means that all initial qubits whose amplitudes of $\ket{L}$ and $\ket{R}$ have opposite phases ($\beta=\pi$) can be accurately estimated, provided these amplitudes are different ($\alpha\neq\pi/2$). This behavior corroborates the symmetry discussed above for $P_E(\alpha,\beta;M\!=\!1)$: qubits on the northern hemisphere of the Bloch sphere have the same escape probability as their counterparts on the southern hemisphere. Only for the qubit $(\alpha,\beta)=(\pi/2,\pi)$, which is completely absorbed by the barrier at $j=1$, is the estimate poor. Similarly, we obtain an excellent estimate for $\beta$ when $\alpha=\pi/2$, except at $\beta=\pi$ and approximately for $\beta<\pi/2$ and $\beta>3\pi/2$, where $F_\beta$ decreases monotonically to zero. In this case, the situation is reversed: the amplitudes of $\ket{L}$ and $\ket{R}$ are equal (qubits on the Bloch sphere’s equator); however, as the relative phase between $\ket{L}$ and $\ket{R}$ approaches zero or $\pi$, the estimate becomes poor. Finite values of $M>1$ are remarkably similar to $M\to\infty$, where we have a distinct behavior in the estimates. In such a case, we have \emph{hot spots} corresponding to two particular qubits ($P_E\approx 0.77$), in which a fine estimate can also be made.

Since both coin-state parameters $(\alpha,\beta)$ are unknown in general (while the coin operator is assumed fixed throughout, e.g., $\rho=1/2$ for the explicit closed forms), it is
natural to consider the Fisher information matrix (per trial)
\begin{equation}
\!\mathcal{F}(\!\alpha,\!\beta;\!M\!)\!=\!\frac{1}{P_E(1\!-\!P_E)}
\begin{bmatrix}
(\partial_{\alpha}\!P_E)^2 & \!\partial_{\alpha}\!P_E\,\partial_{\beta}\!P_E\\
\partial_{\alpha}\!P_E\,\partial_{\beta}\!P_E & \!(\partial_{\beta}\!P_E)^2
\end{bmatrix}\!,
\end{equation}
so for $N$ independent trials we have $N\,\mathcal{F}$. Its determinant and conditioning quantify identifiability of $(\alpha,\beta)$ at a given $M$ and explain the \emph{hot spots} seen in Fig.~\ref{fig.3}. Combining data from multiple boundary placements $\{M_\ell\}$ adds information,
\begin{equation}
\mathcal{F}_{\mathrm{tot}}(\alpha,\beta)=\sum_{\ell}\mathcal{F}(\alpha,\beta;M_\ell), 
\label{joint}
\end{equation}
which mitigates the poor sensitivity of $F_\beta$ at $M=1$ and improves joint estimation via an increased $\det \mathcal{F}_{\mathrm{tot}}$.

Now, instead of performing the quantum walk, we check Eq.~\eqref{Psi0} by direct measurement, whose QFI is given by
\begin{equation}
    H_{\theta} = 4\left[\braket{\partial_\theta\Psi_{\theta}(0)|\partial_\theta\Psi_{\theta}(0)} - \left| \braket{\Psi(0)|\partial_\theta\Psi_{\theta}(0)}\right|^{2} \right].
\end{equation}
For the pure initial coin state of Eq.~\eqref{Psi0}, parametrized by $(\alpha,\beta)$, the single-copy QFI under optimal measurements is $H_{\alpha}=1$ and $H_{\beta}=\sin^{2}\alpha$. Here ``single copy'' means per experimental run; in practice the total Fisher information scales as $N$ for $N$ independent repetitions, yielding a Cramér-Rao scaling $\mathrm{Var}\sim 1/N$.

A convenient performance metric is the (dimensionless) efficiency,
\begin{equation}
\eta_{\theta}(\alpha,\beta;M)=\frac{F_{\theta}(\alpha,\beta;M)}{H_{\theta}}\le 1,
\end{equation}
which quantifies how close absorption readout comes to the ultimate bound within the restricted-readout setting considered here (note that $\eta_{\beta}$ is defined only where $H_{\beta}>0$).

\begin{figure}[h]
\includegraphics[width=\linewidth]{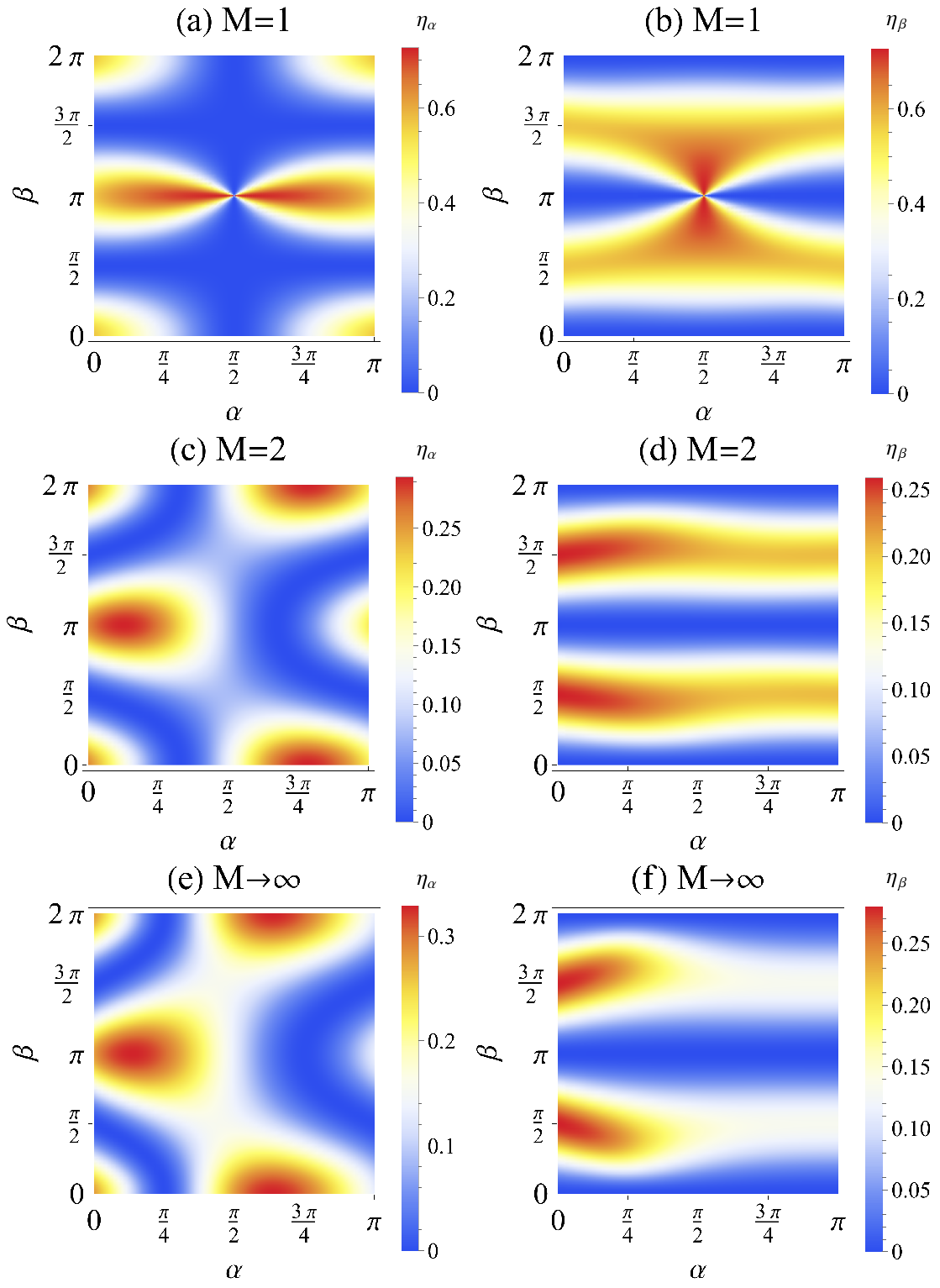}
\caption{\justifying Efficiency of absorption readout relative to the quantum limit. Left column: $\eta_{\alpha}(\alpha,\beta;M)=F_{\alpha}/H_{\alpha}$; right column: $\eta_{\beta}(\alpha,\beta;M)=F_{\beta}/H_{\beta}$. Panels (a), (b) correspond to $M=1$; panels (c), (d) correspond to $M=2$; and panels (e), (f) correspond to $M\to\infty$, respectively. Axes: $\alpha\in[0,\pi]$ (horizontal) and $\beta\in[0,2\pi]$ (vertical). Color encodes per-trial efficiency (dimensionless), capped at the 99th percentile to avoid saturation near singular sets where $P_E(1-P_E)\to 0$. For $M=1$, $\eta_{\alpha}$ is broadly high except near $\alpha=\pi/2$, while $\eta_{\beta}$ is largely suppressed. As $M$ increases, phase-sensitive ``hot spots'' emerge in $\eta_{\beta}$ (notably around $\alpha\approx\pi/2$ and $\beta\approx\pi/2,3\pi/2$), and $\eta_{\alpha}$ becomes more structured. These complementary patterns motivate combining two boundary placements to obtain well-conditioned joint estimation of $(\alpha,\beta)$ and tighter Cramér-Rao bounds at fixed $N$.}
\label{fig:4}
\end{figure}

Figure~\ref{fig:4} displays the efficiencies $\eta_{\alpha}(\alpha,\beta;M)=F_{\alpha}/H_{\alpha}$ (left column) and $\eta_{\beta}(\alpha,\beta;M)=F_{\beta}/H_{\beta}$ (right column) for three boundary placements at $M=1$, $M=2$, and $M\to\infty$. Two robust consequences explain most features: (1) $\eta_{\beta}$ must vanish along $\beta\in\{0,\pi\}$ and near $\alpha\in\{0,\pi\}$ because either $\sin\beta=0$ or $H_{\beta}=\sin^{2}\alpha=0$ and (2) $\eta_{\alpha}$ is suppressed near $\alpha=\pi/2$ unless the interference term $\xi_{3}\cos\beta$ compensates the zero of $\cos\alpha$. In addition, both efficiencies decay close to the singular sets $P_{E}\in\{0,1\}$, where the Bernoulli variance vanishes and the binary measurement carries no information.

For the nearest boundary ($M=1$), $\xi_{1}=\xi_{2}$ and $\xi_{3}>0$, so $F_{\alpha}\propto(\xi_{3}\cos\alpha\cos\beta)^{2}$. The map of $\eta_{\alpha}$ exhibits a broad plateau of high efficiency across most $(\alpha,\beta)$, with a dip centered at $\alpha=\pi/2$. In contrast, $\eta_{\beta}\propto(\sin\alpha\sin\beta)^{2}$ is generically small, showing only weak ridges near $\alpha\simeq\pi/2$ and $\beta\simeq\pi/2,3\pi/2$ (tempered by the Bernoulli factor). Operationally, a single near boundary is highly informative for estimating the population angle $\alpha$, but comparatively less informative for the phase $\beta$.

At moderate distances (e.g., $M=2$), $(\xi_{2}-\xi_{1})\neq 0$ and $\xi_{3}$ is reduced, producing interference between the $\sin\alpha$ and $\cos\alpha\cos\beta$ terms in $F_{\alpha}$. The $\eta_{\alpha}$ landscape remains strong but loses uniformity, while $\eta_{\beta}$ develops localized \emph{hot spots} near $\alpha\approx\pi/2$ and $\beta\approx\pi/2,3\pi/2$, where $|\sin\alpha\sin\beta|$ is maximal and $P_{E}$ stays away from $\{0,1\}$. In the limit ($M\to\infty$), both $(\xi_{2}-\xi_{1})$ and $\xi_{3}$ remain $\mathcal{O}(1)$, sharpening the phase-sensitive islands in $\eta_{\beta}$ and making $\eta_{\alpha}$ more structured. These trends confirm a clean complementarity: near boundaries favor $\alpha$, and far boundaries expose $\beta$.

Because for a single $M$ the per-trial Fisher information matrix is $\mathcal{F}\propto g\,g^{\top}$ [with $g=(\partial_{\alpha}P_{E},\partial_{\beta}P_{E})$], it has rank~1, so joint estimation is informed only along the gradient direction. Consequently, one should combine two (or more) boundary placements, as in Eq. (\ref{joint}), chosen so that the corresponding sensitivity directions are not colinear at the operating point. A practical and effective choice is $M_{1}=1$ (high $\eta_{\alpha}$ across a broad region) together with a moderate or large $M_{2}$ (phase-sensitive \emph{hot spots} in $\eta_{\beta}$). This pairing yields a well-conditioned $\mathcal{F}_{\mathrm{tot}}$, tightening the matrix Cramér-Rao bound at fixed resources $N$.

Remark that absorption readout is not informationally optimal, but it is experimentally natural in platforms where the coin degree of freedom is not directly addressable in arbitrary bases or where full tomography is too costly.  Accordingly, the quantum Fisher information serves here as a benchmark, while the relevant achievable precision under the restricted-access measurement is quantified by the classical Fisher information of the binary (absorbed/escaped) outcome.

\section{Absorption readout under restricted measurement access}\label{sec.5}

We now discuss the operational scope of the present results for integrated-photonic platforms, which already realize discrete-time quantum walks in reconfigurable waveguide arrays and multiport interferometers, with single- and two-photon versions widely demonstrated~\cite{Peruzzo2010,Sansoni2012,Neves2018}. In such devices, characterization often relies on quantum state tomography of path/polarization qubits (or qudits), which demands many measurement settings and phase-stable reconfigurations~\cite{James2001} (see also Ref.~\cite{PhysRevA.82.062308} for another strategy). The present strategy is particularly relevant for integrated-photonic implementations using an on-chip sink. In such architectures, the absorbing boundary may be implemented as a controllable loss channel (e.g., via evanescent coupling to a lossy bus or an integrated detector), and the readout reduces to a binary outcome, escaped versus absorbed. Since the sensitivity patterns for $\alpha$ and $\beta$ are complementary across $M$, combining two boundary placements can yield a well-conditioned Fisher matrix for joint estimation, without requiring reconstruction of the full output state. In this restricted-readout setting, the protocol may reduce measurement-setting and reconfiguration overhead relative to mode-resolved tomography when the goal is low-dimensional parameter estimation rather than full state reconstruction. Therefore, our proposal should not be interpreted as a general replacement for full tomography. Rather, it addresses a different and more limited question: what information about the initial coin qubit remains experimentally accessible when the available readout is restricted to a global binary outcome? In that setting, absorption readout defines a coarse-grained estimation primitive whose information content can be characterized analytically.

\subsection{What absorption readout can and cannot estimate}

Absorption/escape readout is \emph{not} informationally complete. At a fixed boundary placement $M$, the measurement yields a single Bernoulli statistic, $P_E(\alpha,\beta;M)$, and thus provides at most one independent constraint on the underlying state. Consequently, $s$ distinct boundary placements provide at most $s$ independent scalar statistics, and can only identify at most $s$ independent combinations of underlying parameters without further model assumptions. In particular, absorption readout does not enable spatially resolved multimode state tomography of a general quantum-walk output state, nor does it furnish full reconstruction in multiexcitation settings. Its operational value is instead to support low-dimensional parameter estimation and coarse-grained device characterization under restricted measurement access.

In the present single-coin setting, this distinction is especially transparent. For a single boundary position, the Fisher matrix is rank~1 and therefore informs only one local direction in parameter space. By combining two boundary placements with noncolinear sensitivity directions, however, one generically obtains a full-rank total Fisher matrix, enabling joint estimation of $(\alpha,\beta)$ without reconstructing the full output state. The resulting task is therefore parameter estimation under restricted access, not tomography.

\subsection{Resource trade-off and long-time evolution}

An important caveat concerns the role of long-time dynamics. Our analytical expressions are derived for the asymptotic escape probability because absorption probabilities in quantum walks are naturally defined in the long-time limit. This asymptotic formulation should not be understood as implying that long evolution time is an experimentally free resource. On the contrary, in realistic implementations, coherence time, propagation loss, decoherence, detector efficiency, and finite-time convergence all constitute relevant physical constraints. Accordingly, the practical advantage discussed here is \emph{not} a universal reduction of total experimental cost, but rather a potential reduction in \emph{measurement-setting and reconfiguration overhead} for restricted-readout estimation tasks.

\begin{figure*}[t]
\includegraphics[width=0.8\linewidth]{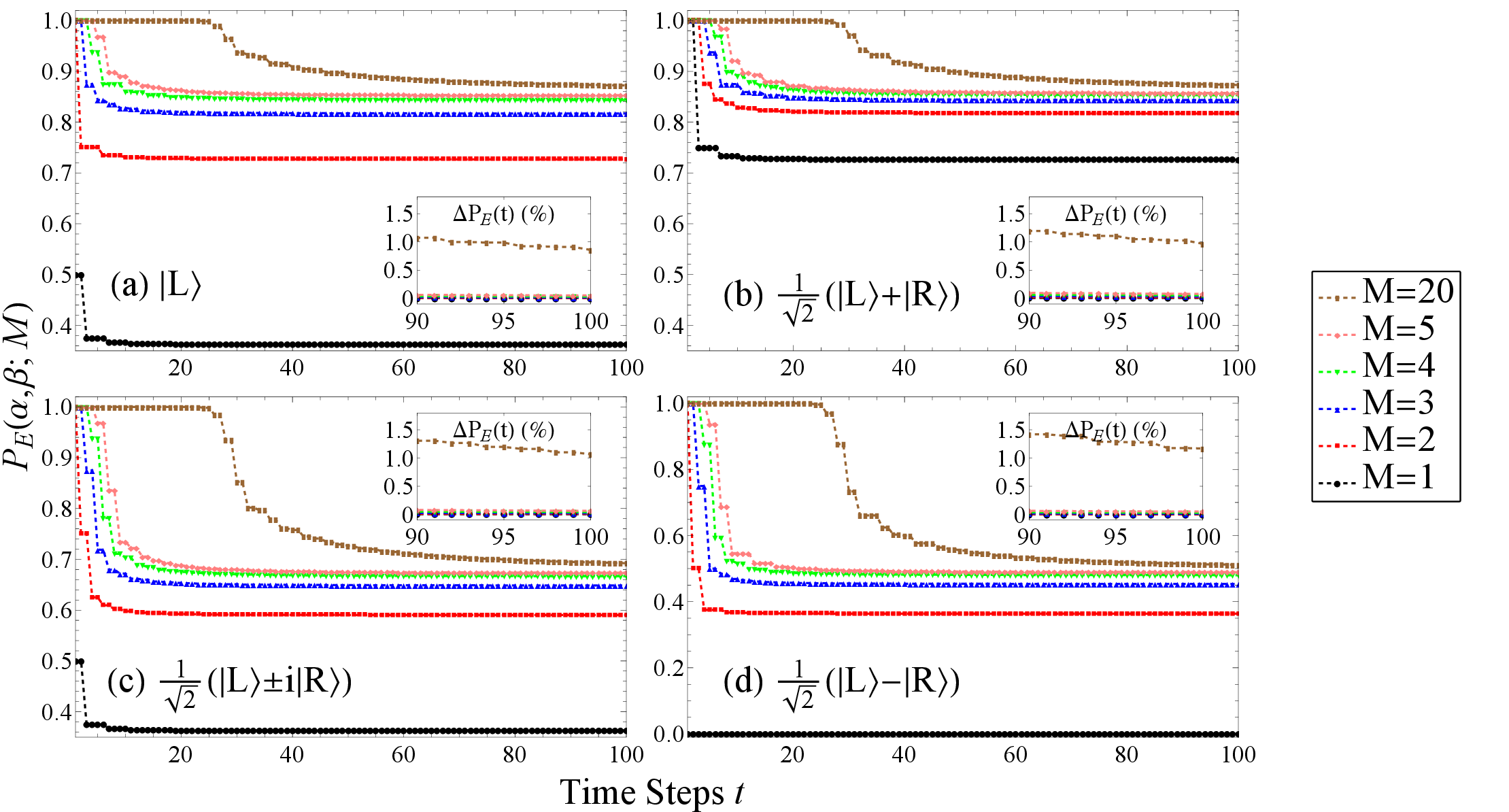}
\caption{\justifying Representative convergence of the finite-time escape probability $P_E(t)$ toward its asymptotic value $P_E(t\to\infty)$ for selected initial coin states and boundary placements. Panels (a)--(d) correspond to different representative choices of $(\alpha,\beta)$. In each panel, the curves correspond to boundary placements $M=1$ (black), $M=2$ (red), $M=3$ (blue), $M=4$ (green), $M=5$ (pink), and $M=20$ (brown), together with the corresponding deviation $\Delta P_E(t)=|P_E(t)-P_E(t\to\infty)|$ shown in the insets. This illustrates that the asymptotic formulas used in the main text correspond to a limiting regime whose experimental accessibility depends on the available evolution time. The convergence rate depends on the operating point in $(\alpha,\beta)$ and on the boundary position $M$, thereby making explicit the trade-off between reduced measurement-setting complexity and the evolution time required for the absorption statistics to approach their asymptotic values. For $M=20$, the deviation $\Delta P_E(t)$ is computed relative to the asymptotic value $P_E(\alpha,\beta;M\to\infty)$ from Eq.~\eqref{PEM}, for comparison.}
\label{fig.5}
\end{figure*}

To illustrate how the asymptotic regime is approached in practice, Fig.~\ref{fig.5} shows representative finite-time escape probabilities $P_E(t)$ for selected initial coin states and boundary positions, together with the deviation $\Delta P_E(t)=|P_E(t)-P_E(t\to\infty)|$. The figure makes explicit that the long-time limit is reached at different rates depending on the operating point in $(\alpha,\beta)$ and on the absorber position $M$. This reinforces that the asymptotic expressions derived above provide a limiting benchmark, while their experimental accessibility is conditioned by the available evolution time.

To make this point explicit, the comparison with tomography should be understood as matched only with respect to measurement access: absorption readout is attractive in architectures where direct projective access to the coin or to individual output modes is unavailable or expensive. A full platform-dependent resource comparison would require incorporating finite-time convergence, losses, decoherence, and other implementation details on the same footing as measurement-setting complexity.

\subsection{Restricted-readout scaling and practical protocol}

With this scope in mind, one may still compare the reconfiguration overhead of different characterization strategies in large-walk devices. Consider a $T$-step one-dimensional walk with support over $K\approx 2T+1$ spatial modes. A coin-qubit tomography that is local and mode resolved requires at least three projective settings (Pauli $X$, $Y$, and $Z$) per mode, i.e., $\mathcal{S}_{\mathrm{tomo}}\sim 3K$ distinct configurations; with $N$ detection events per configuration, the total sample count scales as $N_{\mathrm{tomo}}\sim 3KN$. In contrast, absorption readout at $s$ boundary placements (typically $s=2$) requires only $\mathcal{S}_{\mathrm{abs}}=s$ configurations and $N_{\mathrm{abs}}\sim sN$ samples for comparable statistical confidence along the Fisher-information directions relevant to the restricted estimation problem. For example, $T=50$ implies $K\approx 101$, so that $\mathcal{S}_{\mathrm{tomo}}\sim 303$ whereas $\mathcal{S}_{\mathrm{abs}}=2$. This should not be interpreted as a like-for-like advantage over full tomography, since the two protocols address different inference tasks. Rather, it illustrates the potential reduction in reconfiguration count when the experimental goal is low-dimensional parameter estimation under restricted measurement access.

A practical protocol suggested by our Fisher-information analysis is therefore the following: first, use a near boundary, such as $M_1=1$, to estimate $\alpha$ robustly over broad regions of the Bloch sphere; next, use a moderate or distant boundary, $M_2>1$, to probe the phase-sensitive regions that improve estimation of $\beta$; finally, combine the two datasets through Eq.~(\ref{joint}) to obtain a well-conditioned Fisher matrix for joint estimation. In this way, the analysis provides concrete design guidance even though the measurement remains binary and coarse grained.

The same restricted-readout perspective also clarifies the scope of possible extensions beyond the single-excitation, single-coin setting analyzed here. In systems with multiple excitations or larger effective Hilbert spaces, a single absorption observable would generally probe only collective features of the state, such as averaged polarizations, symmetry sectors, or population imbalances. The associated Fisher information would then quantify which coarse-grained parameter combinations remain identifiable, rather than scaling with the full dimension of the underlying Hilbert space. This interpretation is consistent with the broader use of collective observables in limited-access metrology.

\section{Conclusions}\label{sec.6}

We have derived closed-form expressions for the escape probability of discrete-time quantum walks with a single absorbing boundary, explicitly as functions of the Bloch angles $(\alpha,\beta)$ of the initial coin state and the boundary position $M$. From these expressions, we have obtained analytic Fisher information $F_\alpha$ and $F_\beta$, and benchmarked them against the single-copy quantum Fisher information, $H_\alpha=1$ and $H_\beta=\sin^2\alpha$. The resulting efficiency maps $\eta_\alpha=F_\alpha/H_\alpha$ and $\eta_\beta=F_\beta/H_\beta$ reveal a clear complementarity: near boundaries (e.g., $M=1$) are broadly informative about the polar Bloch angle $\alpha$, whereas moderate or large distances (e.g., $M=2$ and $M\to\infty$) generate phase-sensitive regions informative about $\beta$. Because the per-trial Fisher matrix at fixed $M$ has rank~1, combining two boundary placements can render the information full rank and tighten the joint Cramér-Rao bound while retaining the simplicity of a binary absorption readout.

The main contribution of the present work is to provide an analytical characterization of the information content of absorption readout in a restricted-access setting. Within that scope, absorption measurements furnish a useful primitive for low-dimensional parameter estimation and coarse-grained calibration of quantum-walk devices. Any practical advantage must, however, be understood together with the corresponding experimental tradeoffs: Although binary absorption readout can reduce measurement-setting and reconfiguration overhead, the long-time regime used to define asymptotic escape probabilities remains an experimental resource subject to finite-time convergence, loss, and decoherence constraints. Subject to those limitations, the main contribution of the present work is to show that the information content of absorption readout can be characterized analytically and exploited for low-dimensional parameter estimation under restricted measurement access.

\begin{acknowledgments}
E.P.M.A. thanks J. Longo for the article revision. This work was supported by Conselho Nacional de Desenvolvimento Científico e Tecnológico—CNPq through Grant No. 409673/2022-6. M.C.O. acknowledges the financial support of the National Institute of Science and Technology for Applied Quantum Computing (INCT-CQA) through CNPq Process No. 408884/2024-0 and FAPESP, through the Center for Research and Innovation on Smart and Quantum Materials (CRISQuaM) Process No. 2024/00998-6.
\end{acknowledgments}
\appendix

\section{Spectral decomposition}\label{append:1}

To perform the spectral decomposition, we should use the quantum-walk space spanned by the Fourier-transformed vectors $\ket{k}=\sum_j e^{ikj}\ket{j}$, with $k\in[-\pi,\pi]$. Then, the operator $S_k$ is now diagonal: 
\begin{align}
S_k&(\ket{L}\otimes\ket{k})=e^{ik}\ket{L}\otimes\ket{k}, \nonumber \\
S_k&(\ket{R}\otimes\ket{k})=e^{-ik}\ket{R}\otimes\ket{k}.
\label{Sk}
\end{align}
Therefore, the time-evolution operator $U_k$ now reads
\begin{equation}
\displaystyle
U_k=
\begin{bmatrix}
e^{ik}\sqrt{\rho}     &   e^{ik}\sqrt{1-\rho} \\
e^{-ik}\sqrt{1-\rho}  &  -e^{-ik}\sqrt{\rho}
\end{bmatrix}.
\label{Uk}
\end{equation}
The diagonalization of $U_k$ gives the eigenvalues 
\begin{equation}
\lambda_{k\pm}=\sqrt{\rho}\left(i\sin k\pm\sqrt{\cos^2 k-1+\frac{1}{\rho}}\right)=e^{-i\omega_{k\pm}},
\label{lambda}
\end{equation}
with $\omega_{k+}=-\sin^{-1}(\sqrt{\rho}\sin k)$ and $\omega_{k-}=\pi-\omega_{k+}$. These eigenvalues are associated with the following eigenstates $\Psi_{k\pm}=(A_{k\pm}, B_{k\pm})^T$, such that
\begin{align}
A_{k\pm}&=\frac{1}{\sqrt{2N}}\sqrt{1\pm\frac{\cos k}{\sqrt{1/\rho-\sin^2 k}}}, \nonumber \\
B_{k\pm}&=\pm\frac{e^{-ik}}{\sqrt{2N}}\sqrt{1\mp\frac{\cos k}{\sqrt{1/\rho-\sin^2 k}}},
\label{Ak&Bk}
\end{align}
where $N$ denotes a finite-lattice regularization used to fix plane-wave normalization in quasimomentum space. Physical predictions are obtained after converting the discrete $k$ sum to an integral in the $N\to\infty$ limit, so the final escape probabilities are independent of the particular normalization convention. The term $\pm e^{-ik}$ in Eq.~\eqref{Ak&Bk} is just an arbitrary phase to guarantee that $A_{k\pm}$ are real and positive \cite{bach2004onedimensional}. 

\section{Escape probability}\label{append:2}

\begin{figure}[h]
\includegraphics[width=\linewidth]{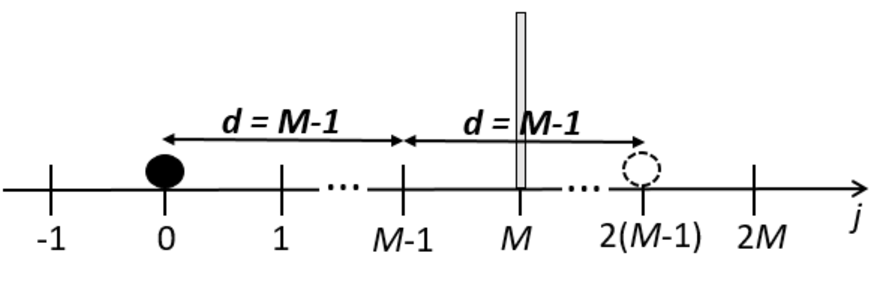}
\caption{\justifying Schematic representation of the \textit{method of images}. The real walker (solid circle) starts at $j=0$, while a mirror walker (open circle) is placed at $j=2(M-1)$. Both walkers are symmetrically positioned around the boundary site, each at a distance $d=M-1$. The superposition of their wave functions ensures that the amplitude $L(M-1,t)$ vanishes at the boundary for all times.}
\label{fig.1}
\end{figure}

As discussed above, when the quantum walker reaches the barrier and no longer returns, the amplitude $L(M-1,t)$ must vanish for all times $t$. However, the eigenstates $\Psi_{k\pm}$ of the system do not satisfy this boundary condition. To enforce $L(M-1,t)=0$, we express the initial condition as a superposition of eigenfunctions such that the contributions from different wave numbers $k$ share the same frequency $\omega$, allowing them to interfere destructively at the barrier. This condition can be conveniently realized through the \textit{method of images}, illustrated in Fig.~\ref{fig.1}. In this approach, two walkers are considered symmetrically positioned with respect to the boundary site $j=M-1$. The real walker is located at $j=0$, and a mirror walker is placed at $j=2(M-1)$. This symmetric construction ensures that the superposition of the real and mirror walkers satisfies the boundary constraint~\cite{bach2004onedimensional,wang2017quantum}.

After employing the method of images, we arrive at the following solution:
\begin{align}
\begin{bmatrix}
L(j,t) \\
R(j,t) \\
\end{bmatrix}
&=\sum_{\substack{k \in \left(-\frac{\pi}{2}, \frac{\pi}{2}\right)}}\!
e^{-i\omega_{k,\pm}t}
\left\{ 
C_{k\pm} 
\begin{bmatrix}
  A_{k\pm} \\ 
  B_{k\pm}
\end{bmatrix}
e^{ikj} \right.\nonumber\\ 
&+
C_{(\pi-k)\pm}
\begin{bmatrix}
  A_{(\pi-k)\pm} \\
  B_{(\pi-k)\pm}
\end{bmatrix}
e^{i(\pi-k)j}
\nonumber\\ 
&+
D_{k\pm} 
\begin{bmatrix}
  A_{k\pm} \\
  B_{k\pm}
\end{bmatrix}
e^{ik[j-2(M-1)]}
\nonumber\\ 
&+\left.
D_{(\pi-k)\pm}
\begin{bmatrix}
  A_{(\pi-k)\pm} \\
  B_{(\pi-k)\pm}
\end{bmatrix}
e^{i(\pi-k)[j-2(M-1)]}
\right\},
\label{REImage}
\end{align}
where $\omega_{k\pm}=\omega_{(\pi-k)\pm}$. Then, $L(M-1,t)=0$ implies 
\begin{equation}
D_{k\pm}=-e^{i\pi(M-1)}C_{(\pi-k)\pm}\frac{A_{(\pi -k)\pm}}{A_{k\pm}}.
\label{Dk}
\end{equation}
Therefore, inside the domain $j<M$ the solution reads
\begin{align}
\begin{bmatrix}
L(j,t)\\
R(j,t)\\
\end{bmatrix}
&= \sum_{\substack{k \in \left(-\frac{\pi}{2}, \frac{\pi}{2}\right)}}
\left\{
\begin{bmatrix}
  A_{k\pm}\\
  B_{k\pm}
\end{bmatrix}
e^{ikj}F_{k\pm} \right.\nonumber\\ 
&+ \left.
\begin{bmatrix}
  A_{(\pi-k)\pm}\\
  B_{(\pi-k)\pm}
\end{bmatrix} 
e^{i(\pi-k)j}G_{k\pm}
\right\},
\label{REfinal}
\end{align}
with
\begin{align}
F_{k\pm} &=C_{k\pm}-C_{(\pi-k)\pm}\frac{A_{(\pi-k)\pm}}{A_{k\pm}}e^{i(\pi-2k)(M-1)},\nonumber\\
G_{k\pm} &=C_{(\pi-k)\pm}-C_{k\pm}\frac{A_{k\pm}}{A_{(\pi-k)\pm}}e^{-i(\pi-2k)(M-1)},
\label{Fk+Gk-}
\end{align}
and taking the local initial condition $(L(0,0), R(0,0))^T$ we conclude \cite{bach2004onedimensional,wang2017quantum} that
\begin{equation}
C_{k\pm}=A_{k\pm}L(0,0)+B^*_{k\pm}R(0,0). 
\label{CAB}
\end{equation}
The solution of Eq.~\eqref{REfinal} consists of a superposition of wave functions with components going to the left and right. The components on the left are the only ones that will be on the physical domain $j<M$ in the long-time limit. These components survive the barrier and compose the escape probability of the particle,
\begin{equation}
P_E(M)=\sum_{k\in\left(-\frac{\pi}{2},\frac{\pi}{2}\right)} |F_{k+}(M)|^2+|G_{k-}(M)|^2. 
\label{PE}
\end{equation}
Replacing  Eqs.~\eqref{Ak&Bk} and \eqref{CAB} into Eq.~\eqref{Fk+Gk-}, we arrive at $G_{k-}=F^*_{k+}$ to rewrite the escape probability as
\begin{equation}
P_E(M)=\sum_{k\in\left(-\frac{\pi}{2},\frac{\pi}{2}\right)} 2|F_{k+}(M)|^2,
\label{PE2}
\end{equation}
with
\begin{align}
F_{k+}(M) &= \frac{L(0,0)}{A_{k+}}
\left(A^2_{k+}\!-\!e^{i(\pi\!-\!2k)(M\!-\!1)}A^2_{k-}\right)\nonumber \\  
&+\!R(0,0)A_{k-}\left(e^{ik}\!+\!e^{i(\pi\!-\!2k)(M\!-\!1)}e^{-ik}\right). 
\label{Fk+}
\end{align}

After converting the sum of Eq.~\eqref{PE2} to an integral, we solve the resulting integrals over $k$ assuming a Hadamard walk ($\rho=1/2$). We follow the steps of Ref. \cite{bach2004onedimensional}. Our interest relies only on assessing the escape probability of a walker positioned closest and farthest possible to the absorption barrier. Then, we obtain the escape probability $P_E(M)$ toward $j\rightarrow -\infty$ for the cases with $M=1$ to $5$ and $M\rightarrow +\infty$ (see Table~\ref{tab.1}). Note that the boundary imposed by $L(M-1,t)=0$ restricts the initial qubit on $j=0$ to have $L(0,0)=0$ with a barrier on $M=1$. This restriction could be easily managed by evolving the state for one time step by hand \cite{bach2004onedimensional}. For instance, taking an initial state $(\cos(\alpha/2)\ket{L}+e^{i\beta}\sin(\alpha/2)\ket{R})\otimes\ket{0}$, after one step with a neighbor barrier, it evolves to the unabsorbed state $(\cos(\alpha/2)+e^{i\beta}\sin(\alpha/2))\ket{L}\otimes\ket{-1}$. The escape probability of such a state is equivalent to $P_E(2)_L$, where the subindex $L$ indicates a walk starting from a qubit $\ket{L}$; we have then $P_{E}(1)=(1+\sin\alpha)^2 P_{E}(2)_L/2$.

\end{document}